\documentclass[a4paper,12pt]{article}

\usepackage{amsmath}
\usepackage[final]{graphicx}% Include figure files
\usepackage{bm}% bold math
\usepackage{amssymb}
\usepackage{amssymb}
\newcounter{comment}

{\refstepcounter{comment}%
\begin{quote}
\ttfamily\small$\blacksquare$ \textbf{\underline{Comment} $\sharp$\thecomment:}}%
{\end{quote}}

\begin{document}
\hfill
%\begin{minipage}{20ex}\small
%ZAGREB-ZTF-09-02\\
%\end{minipage}

\begin{center}
\baselineskip=2\baselineskip
\textbf{\LARGE{Enhancement of $h \rightarrow \gamma \gamma$\\
by seesaw-motivated exotic scalars}}\\[6ex]
\baselineskip=0.5\baselineskip

{\large
Ivica~Picek
%$^{a,}$
%\footnote{picek@phy.hr;
% corresponding author
%}
and
Branimir~Radov\v{c}i\'c
%$^{a,}$
%\footnote{bradov@phy.hr}
}\\[4ex]
\begin{flushleft}
\it
%$^{a}$
Department of Physics, Faculty of Science, University of Zagreb,
 P.O.B. 331, HR-10002 Zagreb, Croatia\\[3ex]
\end{flushleft}
\today \\[5ex]
\end{center}

\begin{abstract}

We examine the role of seesaw-motivated exotic scalars in loop-mediated Higgs decays. We consider a simple TeV-scale seesaw model built upon the fermionic quintuplet mediator in conjunction with the scalar quadruplet, where we examine portions of the model parameter space for which the contributions of charged components of the scalar quadruplet significantly increase the $h \rightarrow \gamma \gamma$ decay rate. The most significant change in the diphoton width comes from a doubly charged scalar $\Phi^{--}$ which should be the lightest component in the scalar quadruplet. In the part of the parameter space where the $h \rightarrow \gamma \gamma$ decay width is enhanced by a factor 1.25 -- 2 there is a mild suppression of the $h \rightarrow Z \gamma$ decay width by a factor 0.9 -- 0.7.

\end{abstract}
\vspace*{2 ex}

\begin{flushleft}
\small
\emph{PACS}:
14.80.Fd; 12.60.Fr; 14.60.Pq
\\
\emph{Keywords}:
Charged Higgs bosons; Extensions of Higgs sector; Neutrino mass
\end{flushleft}

\clearpage

\section{Introduction}

The recently discovered resonance with mass $m_h \simeq 125 - 126$ GeV ~\cite{:2012gk,:2012gu} strongly resembles the Standard Model (SM) Higgs boson. There is a hint that
the loop-induced $h \rightarrow \gamma \gamma$ event rate \cite{Chatrchyan:2012tw,ATLAS:2012ad} deviates, modulo QCD uncertainties \cite{Baglio:2012et}, by a factor 1.5 - 2 \cite{Giardino:2012dp,Corbett:2012dm} from its SM value,
\begin{eqnarray}\label{expt}
{\left[\sigma(gg\to h) \times {\rm BR} (h\to \gamma\gamma) \right]_{\rm LHC}\over \left[\sigma(gg\to h)\times {\rm BR} (h\to \gamma \gamma)\right]_{\rm SM }} = 1.71 \pm 0.33  \ .
\end{eqnarray}
This indicates the existence of additional charged particle(s) on which the tree-level Higgs decay modes are much less sensitive. The history of the ``prediscovery'' of charmed and top quarks through the loop amplitudes may be repeated in the scalar sector.

On the other hand, if this enhancement in $h\rightarrow \gamma \gamma$ disappears with a larger integrated luminosity, it will still constrain the parameter space of various extensions of the SM containing new charged states which should affect this loop amplitude. In this spirit there is a number of theoretical attempts to match the indicated discrepancy by the effects of an extended
Higgs sector \cite{Carena:2012xa,Chang:2012ta,Chiang:2012qz,Dorsner:2012pp}.

Notable,  the $h \rightarrow \gamma \gamma$ decay rate is  sensitive on the doubly charged scalar fields such as those existing in Higgs triplet model of neutrino mass generation.
These states have been in focus of both the direct searches at the LHC ~\cite{Chatrchyan:2012ya,Aad:2012cg} and  of theoretical investigations
 ~\cite{Melfo:2011nx,Arhrib:2011vc,Kanemura:2012rs,Akeroyd:2012ms,Chun:2012jw}. Here we go a step further by studying the effects of singly and doubly charged components of
a scalar quadruplet contained in our recent TeV-scale seesaw model for neutrino masses~\cite{Kumericki:2012bh}.

The Letter is organized as follows. In Section 2 we briefly review the TeV-scale seesaw model at hand. A detailed study of the scalar potential is presented in Section 3. We investigate the effects of exotic scalars on the loop-level Higgs decays in Sections 4 and 5.
Possible direct bounds on exotic scalars are discussed in Section 6. The conclusion is presented in Section 7.

\section{Fermionic quintuplet seesaw model}

A simple and predictive TeV-scale seesaw model~\cite{Kumericki:2012bh} under consideration belongs to a class
of beyond dimension-five seesaw models~\cite{Babu:2009aq,Picek:2009is,Kumericki:2011hf} which, on account of introducing higher iso-multiplets,
lower the originally high seesaw scale to the TeV scale accessible at the LHC. In order to produce the seesaw diagram in the simple model at hand, the fermionic hypercharge-zero quintuplets  $\Sigma_R=(\Sigma_R^{++},\Sigma_R^{+},\Sigma_R^{0},\Sigma_R^{-},\Sigma_R^{--})$ transforming as $(1,5,0)$ under the SM gauge group have to be accompanied by a scalar quadruplet $\Phi=(\Phi^{+},\Phi^{0},\Phi^{-},\Phi^{--})$ transforming as $(1,4,-1)$.

The gauge invariant and renormalizable Lagrangian involving these new fields reads
\begin{equation}\label{lagrangian}
   \mathcal{L} = \overline{\Sigma_R} i \gamma^\mu D_\mu \Sigma_R + (D^\mu \Phi)^\dag (D_\mu \Phi) -
   \big(\overline{L_L} Y \Phi  \Sigma_R + {1 \over 2} \overline{(\Sigma_R)^C} M \Sigma_R + \mathrm{H.c.}\big)
- V(H,\Phi)  \ ,
\end{equation}
where the scalar potential contains the mass and the quartic terms for the doublet $H$ and the quadruplet $\Phi$ fields
\begin{eqnarray}\label{scalarpot}
\nonumber  V(H,\Phi) &=& -\mu_H^2 H^\dagger H + \mu_\Phi^2 \Phi^\dagger \Phi + \lambda_1 \big( H^\dagger H \big)^2 + \lambda_2 H^\dagger H \Phi^\dagger \Phi + \lambda_3 H^* H \Phi^* \Phi \\
\nonumber   &+& \big( \lambda_4 H^* H H \Phi+ \mathrm{H.c.} \big) + \big( \lambda_5 H H \Phi \Phi + \mathrm{H.c.} \big) + \big( \lambda_6 H \Phi^* \Phi \Phi+ \mathrm{H.c.} \big)\\
            &+& \lambda_7 \big( \Phi^\dagger \Phi \big)^2 + \lambda_8 \Phi^* \Phi \Phi^* \Phi  \ .
\end{eqnarray}
In the tensor notation the terms in Eqs.~(\ref{lagrangian}) and~(\ref{scalarpot}) read~\cite{Kumericki:2012bh}
\begin{eqnarray}\label{tensor}
\overline{L_L} \Phi \Sigma_R = \overline{L_L}^i \Phi_{jkl} \Sigma_{Rij'k'l'} \epsilon^{jj'}\epsilon^{k k'}\epsilon^{l l'} &,& \overline{(\Sigma_R)^C} \Sigma_R = \overline{(\Sigma_R)^C}_{ijkl} \Sigma_{Ri'j'k'l'}
\epsilon^{ii'}\epsilon^{jj'}\epsilon^{kk'}\epsilon^{ll'}\ ,\nonumber\\
H^* H \Phi^* \Phi = H^{*i} H_j \Phi^{*jkl} \Phi_{ikl} &,& H^* H H \Phi = H^{*i} H_j H_k \Phi_{ij'k'} \epsilon^{jj'}\epsilon^{kk'}   \ ,\nonumber\\
H H \Phi \Phi =H_i \Phi_{jkl} H_{i'} \Phi_{j'k'l'} \epsilon^{ij}\epsilon^{i'j'}\epsilon^{kk'}\epsilon^{ll'}  &,& H \Phi^* \Phi \Phi = H_i \Phi^{*ijk} \Phi_{jlm} \Phi_{kl'm'} \epsilon^{ll'}\epsilon^{mm'} \ ,\nonumber\\
\Phi^* \Phi \Phi^* \Phi = \Phi^{*ijk} \Phi_{i'jk} \Phi^{*i'j'k'} \Phi_{ij'k'} &.&
\end{eqnarray}

\begin{figure}
\centerline{\includegraphics[scale=1.20]{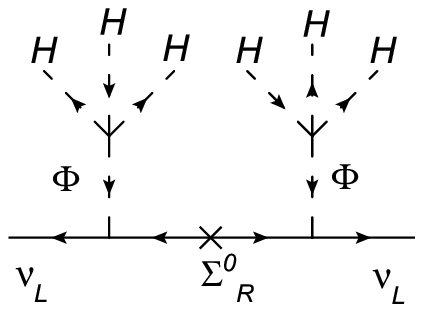}\hspace{1cm}\includegraphics[scale=1.20]{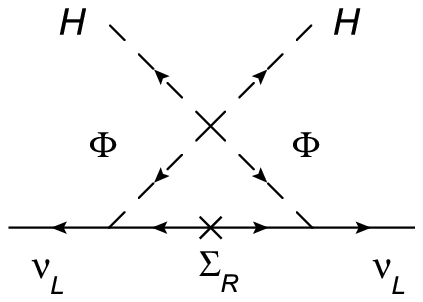}}
\caption{\small Tree-level dimension-nine operator and one-loop dimension-five operator diagrams relevant for generating the light neutrino masses.}
\label{diagrams}
\end{figure}

The light neutrino masses arise through tree-level dimension-nine operator and one-loop dimension-five operator. The diagrams displayed in Fig.~\ref{diagrams} lead to two contributions to light neutrino masses
\begin{eqnarray}\label{mnu}
  (m_\nu)_{ij} &=& (m_\nu)_{ij}^{tree}+(m_\nu)_{ij}^{loop} \nonumber\\
               &=& \frac{-1}{6} (\lambda^*_4)^2 \frac{v_H^6}{\mu_\Phi^4} \sum_k {Y_{ik} Y_{jk} \over M_k}
+ {-5 \lambda_5^* v_H^2 \over 24 \pi^{2}}
\sum_k {Y_{ik} Y_{jk} M_k \over m_\Phi^{2} - M_k^{2}} \left[
1 - {M_k^{2} \over m_\Phi^{2}-M_k^{2}} \ln {m_\Phi^{2} \over M_k^{2}}  \right] \ , \nonumber\\
\end{eqnarray}
determined by quartic couplings $\lambda_4$ and $\lambda_5$, respectively. Let us remind~\cite{Kumericki:2012bh} that in the regime of comparable and light ($\sim$ 200 GeV) masses
for exotic fermions and scalars the tree contribution prevails, while the loop contribution dominates for heavy masses ($\sim$ 500 GeV) and remains a sole contribution
if additional discrete $Z_2$ symmetry forbids the $\lambda_4$ term in Eq.~(\ref{scalarpot}).

\section{Scalar potential}

After the electroweak symmetry breaking (EWSB) the neutral components of the scalar fields acquire a vacuum expectation value and read
\begin{equation}
    H^0=\frac{1}{\sqrt{2}}(v_H+h^0+i\chi)\ , \ \Phi^0=\frac{1}{\sqrt{2}} (v_\Phi+\varphi^0+i\eta) \ .
\end{equation}
If all couplings in the scalar potential are real the conditions for the minimum of the potential
\begin{equation}\label{condition}
    \frac{\partial V_0(v_H,v_\Phi)}{\partial v_H}=0 \ , \ \frac{\partial V_0(v_H,v_\Phi)}{\partial v_\Phi}=0
\end{equation}
give
\begin{eqnarray}
  \mu_H^2 &=& \lambda_1 v_H^2 + (\frac{1}{2}\lambda_2+\frac{1}{6}\lambda_3-\frac{2}{3}\lambda_5)v_\Phi^2 + \frac{\sqrt{3}}{2} \lambda_4 v_H v_\Phi - \frac{\sqrt{3}}{9} \lambda_6 \frac{v_\Phi^3}{v_H}\ ,\\
  \mu_\Phi^2 &=& -(\frac{1}{2}\lambda_2+\frac{1}{6}\lambda_3-\frac{2}{3}\lambda_5)v_H^2 - \frac{\sqrt{3}}{6} \lambda_4 \frac{v_H^3}{v_\Phi} + \frac{\sqrt{3}}{3} \lambda_6 v_H v_\Phi - (\lambda_7+\frac{5}{9}\lambda_8) v_\Phi^2\ .\nonumber\\
\end{eqnarray}
The electroweak $\rho$ parameter is changed by $v_\Phi$ from the unit value to $\rho\simeq1+6 v_\Phi^2/v_H^2$ constraining the ratio $v_\Phi/v_H$ to be smaller then 0.015.

The mixing between singly charged and between neutral components of $H$ and $\Phi$ multiplets will occur after the EWSB. Let us illustrate this on the $h^0 - \varphi^0$ mass terms
\begin{eqnarray}
\mathcal{L}_{h^0 \varphi^0} = -\frac{1}{2} \left( -\mu_H^2 + 3\lambda_1 v_H^2 +(\frac{1}{2}\lambda_2+\frac{1}{6}\lambda_3-\frac{2}{3}\lambda_5) v_\Phi^2 +\sqrt{3}\lambda_4 v_H v_\Phi \right) h^0 h^0  \nonumber \\
-\left( (\lambda_2+\frac{1}{3}\lambda_3- \frac{4}{3}\lambda_5) v_H v_\Phi + \frac{\sqrt{3}}{2}\lambda_4 v_H^2 - \frac{1}{\sqrt{3}}\lambda_6 v_\Phi^2\right) h^0 \varphi^0  \nonumber \\
-\frac{1}{2} \left( \mu_\Phi^2 +(\frac{1}{2}\lambda_2+\frac{1}{6}\lambda_3-\frac{2}{3}\lambda_5) v_H^2 +\frac{2}{\sqrt{3}}\lambda_6 v_H v_\Phi +(3\lambda_7 + \frac{5}{3}\lambda_8) v_\Phi^2    \right) \varphi^0 \varphi^0 \ .\nonumber\\
\end{eqnarray}
The mass term that mixes $h^0$ and $\varphi^0$ is small because it is proportional either to $v_\Phi$ from $\lambda_{2,3,6}$ terms in Eq.~(\ref{scalarpot}), or to small lepton number violating couplings $\lambda_{4,5}$ dictated to be small from considerations of neutrino masses. Therefore we neglect the mixing between the components of $H$ and $\Phi$ multiplets and we neglect the terms of higher order in $v_\Phi/v_H$ whenever possible.

In the approximations given above, the masses of charged components of the quadruplet $\Phi$ are
\begin{eqnarray}\label{spectrum}
  m^2(\Phi^{+}) &=& \mu_\Phi^2 + \frac{1}{2}\lambda_2v_H^2 \ , \nonumber\\
  m^2(\Phi^{-}) &=& \mu_\Phi^2 + \frac{1}{2}\lambda_2v_H^2 + \frac{1}{3}\lambda_3v_H^2\ , \\
  m^2(\Phi^{--}) &=& \mu_\Phi^2 + \frac{1}{2}\lambda_2v_H^2 + \frac{1}{2}\lambda_3v_H^2\ . \nonumber
\end{eqnarray}
Their couplings to the Higgs boson relevant for the $h \rightarrow \gamma \gamma$ decay are
\begin{equation}
    - \mathcal{L} = c_{\Phi^{+}} v_H h^0 \Phi^{+*} \Phi^{+} + c_{\Phi^{-}} v_H h^0 \Phi^{-*} \Phi^{-} + c_{\Phi^{--}} v_H h^0 \Phi^{--*} \Phi^{--} \ ,
\end{equation}
where the newly introduced couplings
\begin{equation}\label{couplings}
    c_{\Phi^{+}}= \lambda_2 \ ,\ c_{\Phi^{-}}= \lambda_2 + \frac{2}{3} \lambda_3 \ , \ c_{\Phi^{--}}= \lambda_2 + \lambda_3 \ ,
\end{equation}
are expressed in terms of the quartic couplings $\lambda_2$ and $\lambda_3$ which, for simplicity, we assume to be equal in the following.

\section{Higgs diphoton-decay width}

In our model in addition to the dominant SM contributions from the $W$ boson and top quark loops to the $h \rightarrow \gamma \gamma$ decay rate only the charged scalars contribute substantially. 
Our exotic scalars are colorless so that the Higgs boson production through gluon fusion at the LHC is unaffected.  Also, since the Higgs boson total decay width is only marginally affected,
a dominant change in $h \rightarrow \gamma \gamma$ event rate comes from the change in Higgs boson diphoton partial decay width.
The analytic expression for the diphoton $h \rightarrow \gamma \gamma$ partial width reads ~\cite{Carena:2012xa,Ellis:1975ap,Shifman:1979eb,Djouadi:2005gi}
\begin{equation}
\label{W-t-S-loop}
\Gamma(h\to \gamma \gamma)=\frac{\alpha^2 m_h^3}{256 \pi^3 v_H^2}\left|A_1(\tau_W)+ N_c Q_t^2  A_{1/2}(\tau_t)
+  N_{c,S} Q_S^2 \frac{c_S}{2} \frac{v_H^2}{m_S^2} A_0(\tau_S)\right |^2 \ ,
\end{equation}
where the three contributions corresponding to $\tau_i\equiv 4m_i^2/m_h^2$ ($i=W, t, S$) refer to  spin-1 ($W$ boson), spin-1/2 (top quark) and  charged spin-0 particles in the loop.
The electric charges for fermions and scalars,  $Q_t=+2/3$ and $Q_S$, are given in units of $|e|$, and their respective number of colors are $N_c=3$ and $N_{c,S}=1$ for color singlet scalars under consideration.
The loop functions  for spin-1, spin-1/2, and spin-0 particles are given as
\begin{eqnarray}
 \label{W-loop}
A_1(x)&=& -x^2\left[2x^{-2}+3x^{-1}+3(2x^{-1}-1)f(x^{-1})\right]\ ,\\
\label{t-loop}
A_{1/2}(x) &=& 2  \, x^2 \left[x^{-1}+ (x^{-1}-1)f(x^{-1})\right] \ ,\\
 \label{S-loop}
A_0(x) &=& -x^2 \left[x^{-1}-f(x^{-1})\right]  \ ,
\end{eqnarray}
 where, for a Higgs mass below the kinematic threshold of the loop particle $m_h < 2 \; m_{\rm loop}$, we have
\begin{equation}
 f(x) = \arcsin^2 \sqrt{x} \ .
\end{equation}
\begin{figure}
\centerline{\includegraphics[scale=0.65]{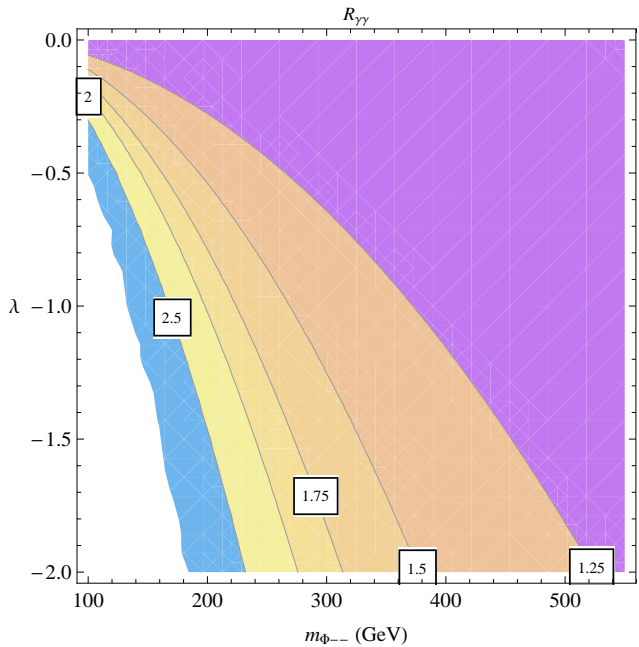}}
\caption{\small Enhancement factor $R_{\gamma\gamma}$ =(2.5, 2, 1.75, 1.5, 1.25) for the $h \rightarrow \gamma \gamma$ branching ratio in dependence on scalar coupling $\lambda =\lambda_2=\lambda_3$
and the mass of the lightest charged component of the $\Phi$ multiplet, $m(\Phi^{--})$.}
\label{rgamma}
\end{figure}
We now look at the possible enhancement of the $h \rightarrow \gamma \gamma$ decay rate through extra contributions with charged components of scalar quadruplet $\Phi$ running in the loops. Following \cite{Carena:2012xa} we define the enhancement factor with respect to the SM decay width
\begin{equation}\label{Rgamgam}
    R_{\gamma\gamma} = \left| 1+  \sum_{S=\Phi^{+},\Phi^{-},\Phi^{--}} Q_S^2 \frac{c_S}{2} \frac{v_H^2}{m_{S}^2}\frac{A_0(\tau_{S})}{ A_1(\tau_W)+ N_c Q_t^2 \, A_{1/2}(\tau_t)}\right|^2 \ .
\end{equation}

In order to get an enhancement of the $h \rightarrow \gamma \gamma$ decay rate, the contribution of the new charged scalars has to interfere constructively with the dominant SM contribution of the $W$ boson. Since this requires negative couplings $c_S$ in Eq.~(\ref{W-t-S-loop}), we assume $\lambda_{2,3}<0$. The biggest effect on the $h \rightarrow \gamma \gamma$ decay rate comes from $\Phi^{--}$ for two reasons. First, it has a charge $|Q_{\Phi^{--}}|=2$ and second, for $\lambda_3<0$ it is the lightest of the charged $\Phi$ components. We will not consider $\lambda_{2,3}>0$ case for which to get a significant enhancement in $R_{\gamma\gamma}$ the charged scalars should be too light.

In Fig.~\ref{rgamma} we plot the enhancement factor $R_{\gamma\gamma}$ as a function of the scalar couplings $\lambda_{2,3}$ and the mass of the lightest charged scalar $m(\Phi^{--})$. For simplicity we assume $\lambda_2=\lambda_3=\lambda$, and in order to keep the stability of the scalar potential, we restrict ourselves to the interval $\lambda \in [-2,0]$. In this interval the value $R_{\gamma\gamma}=2$ can be achieved up to $m(\Phi^{--})=280$ GeV and the value $R_{\gamma\gamma}=1.25$ up to $m(\Phi^{--})=520$ GeV.

\section{Higgs to Z-photon decay width}

Another loop-mediated Higgs decay sensitive to new charged particles is $h \rightarrow Z \gamma$. It is sensitive both on the charge and weak isospin of the particles in the loop. After adding a contribution of a new scalar to the SM contributions from the $W$ boson and top quark the decay rate is given by \cite{Carena:2012xa}
\begin{equation}
    \Gamma(h\to Z\gamma)= \frac{\alpha^2 m_h^3}{128 \pi^3 v_H^2 \sin^2 \theta_w} \left(1-\frac{m_Z^2}{m_h^2}\right)^3 \left| {\cal A}_{SM} + {\cal A}_S\right|^2 \ ,
\end{equation}
where
\begin{equation}
    {\cal A}_{SM} = \cos \theta_w A_1(\tau_W,\sigma_W) + N_c \frac{Q_t (2T_3^{(t)}-4 Q_t \sin^2 \theta_w)}{\cos \theta_w} A_{1/2}(\tau_t,\sigma_t) \ ,
\end{equation}
\begin{equation}
    {\cal A}_S = \frac{v \sin \theta_w}{2}\ \frac{c_S v_H}{m_S^2}\ Q_S \ \frac{T_3^{(S)}-Q_S \sin^2 \theta_w}{\cos \theta_w \sin \theta_w} A_0(\tau_S,\sigma_S)\ .
\end{equation}
Here $\sigma_i=4m_i^2/m_Z^2$ and $T_3^{(t)}=1/2$ and $T_3^{(S)}$ correspond to the weak isospin of the top quark and the new scalar. The loop functions are given by
\begin{eqnarray}
A_1(x,y)&=& 4 (3-\tan^2\theta_w) I_2(x,y)+ \left[ (1+2x^{-1}) \tan^2\theta_w - (5+2x^{-1})\right] I_1(x,y) \ , \nonumber\\
 A_{1/2}(x,y)& =& I_1(x,y)-I_2(x,y) \ , \nonumber\\
A_{0}(x,y)& =& I_1(x,y) \ ,
\end{eqnarray}
where
\begin{eqnarray}
 I_1(x,y) &=& \frac{x y}{2(x-y)} + \frac{x^2 y^2}{2(x-y)^2}[ f(x^{-1})-f(y^{-1})] + \frac{x^2 y}{(x-y)^2}[g(x^{-1})-g(y^{-1})] \ ,\nonumber\\
 I_2(x,y) &=& - \frac{x y}{2(x-y)} [ f(x^{-1})-f(y^{-1})]  \ .
\end{eqnarray}
For the Higgs mass below the kinematic threshold of the loop particle, $m_h < 2 \; m_{\rm loop}$, we have
\begin{equation}
     g(x) = \sqrt{x^{-1} -1} \arcsin \sqrt{x} \ .
\end{equation}
The additional contributions in a model with scalar quadruplet lead to the modification factor $R_{Z\gamma}$ for the $h \rightarrow Z \gamma$ decay rate with respect to the SM decay width
\begin{equation}
    R_{Z\gamma} = \left| 1+  \sum_{S=\Phi^{+},\Phi^{-},\Phi^{--}} \frac{{\cal A}_S}{{\cal A}_{SM}}\right|^2 \ .
\end{equation}

\begin{figure}[h]
\centerline{\includegraphics[scale=0.65]{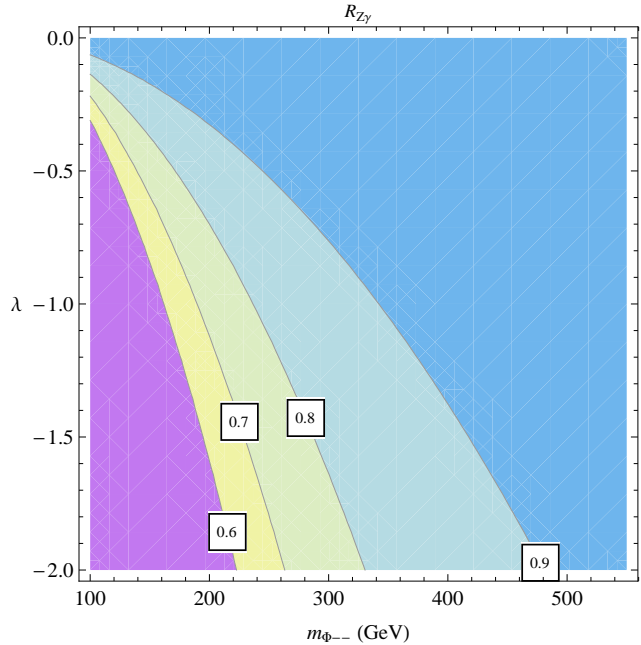}}
\caption{\small Modification factor $R_{Z\gamma}$ =(0.6, 0.7, 0.8, 0.9) for the $h \rightarrow Z \gamma$ branching ratio in dependence on the scalar coupling  $\lambda =\lambda_2=\lambda_3$
and the mass of the lightest charged component of the $\Phi$ multiplet, $m(\Phi^{--})$.}
\label{rZg}
\end{figure}

In Fig.~\ref{rZg} we plot the modification factor $R_{Z\gamma}$ as a function of the scalar couplings $\lambda_{2,3}$ (assuming $\lambda_2=\lambda_3=\lambda$) and the mass of the lightest charged scalar $m(\Phi^{--})$. We obtain a moderate suppression of the $h \rightarrow Z \gamma$ decay rate in the region of the parameter space where the $h \rightarrow \gamma \gamma$ decay rate is enhanced. In the chosen interval $\lambda \in [-2,0]$ the factor $R_{Z\gamma}=0.6$ is achieved up to $m(\Phi^{--})=220$ GeV and $R_{Z\gamma}=0.9$ up to $m(\Phi^{--})=480$ GeV.

\section{Direct vs. virtual bounds on new scalars}

The measurement of the $h \rightarrow \gamma \gamma$ rare decay mode may mark a spectacular onset of a virtual physics at the LHC. The interaction with virtual particles
is essential both for the production and the decay factor in Eq.~(\ref{expt}). It is in order to discuss the features of additional charged scalars, which determine their
possible appearance both in the Higgs boson loop decays under consideration and in their direct searches.

Now, an essential point is a lack of Yukawa couplings of the scalar quadruplet to a pair of SM fermions. The lack of these Yukawa couplings means that a doubly charged $\Phi^{--}$ does not decay to a pair of charged SM leptons, which has decisive repercussions. The direct searches for doubly charged scalars through resonance in the invariant mass of a pair of charged leptons at the LHC ~\cite{Chatrchyan:2012ya,Aad:2012cg} leave the quadruplet scalar states unconstrained. The strongest bound in the range (383 GeV, 408 GeV) set in the inclusive search for the CMS benchmark points depends on $\Phi^{--}$ production analysis, and counts all possible final state lepton pairs in BR($\Phi^{--}\rightarrow l^-_i l^-_j $), $i, j = e, \mu, \tau$.
Of course, $\Phi^{--}$ can decay also to $W^- W^-$ final state, but this channel has not been measured yet and doesn't
constrain the quadruplet scalar  mass. 
 
In comparison, the studies of $h \rightarrow \gamma \gamma$ enhancement by the doubly charged scalar from the Higgs triplet model of neutrino mass generation~\cite{Melfo:2011nx,Arhrib:2011vc,Kanemura:2012rs,Akeroyd:2012ms} also lead to their rather small masses, which already seem to be excluded by the above mentioned direct searches at the LHC ~\cite{Chatrchyan:2012ya,Aad:2012cg}.
In case that the decay of the doubly charged component of the triplet scalar
to $W^- W^-$  were the dominant one, it would invalidate the discussed CMS search and no limit on the triplet mass would be placed either.

\section{Conclusion}

It is conceivable that the first sign of new physics observed at the LHC may stem from loop effects of new particles additional to the SM content. We take under scrutiny
particular extensions of the SM Higgs sector aimed to explain the small masses of neutrinos, which simultaneously provide doubly charged scalars that significantly
affect the loop amplitude for the $h \rightarrow \gamma \gamma$ decay. In view of the upper mentioned exclusion of the doubly charged scalars from the Higgs triplet model of
neutrino masses,
we find it timely to explore the effect of doubly charged component of the scalar quadruplet existing in our recent simple TeV-scale seesaw model
for neutrino masses~\cite{Kumericki:2012bh}. The results which we obtain show that it is easier to obtain an enhancement of the $h \rightarrow \gamma \gamma$ decay rate in our model than in the Higgs triplet model.
The Higgs quadruplet is devoid of the Yukawa couplings to a pair of leptons which play a decisive role in existing LHC exclusion bounds for doubly charged scalars. Therefore our model
in the sub-TeV mass region, where a significant enhancement in $R_{\gamma\gamma}$ can be obtained, is not excluded.
Also, there is a remarkable feature of achieving the biggest effect on $h \rightarrow \gamma \gamma$ decay rate from the doubly charged $\Phi^{--}$ state at hand. The essential $\lambda_3<0$ coupling ensures that $\Phi^{--}$ is the lightest among the charged $\Phi$ components and that it interferes constructively with the dominant SM contribution of the $W$ boson.
Concerning the monitoring $h \rightarrow Z \gamma$ decay, the effect is smaller and opposite. There is a moderate suppression of the $h \rightarrow Z \gamma$ decay
rate in the region of the parameter space where the $h \rightarrow \gamma \gamma$ decay rate is enhanced.
Finally, the imposed role on the scalar quadruplet in explaining the smallness  of neutrino masses is decisive in keeping the relevant couplings from the
scalar potential small so that the mixing between components of $H$ and $\Phi$ multiplets is negligible. Accordingly, the state announced at the LHC~\cite{:2012gk,:2012gu}
corresponds in our setup to the SM Higgs boson.

\subsubsection*{Acknowledgements}

We thank Kre\v{s}imir~Kumeri\v{c}ki for useful discussions.
This work is supported by the Croatian Ministry  of Science, Education and Sports under Contract No. 119-0982930-1016.

\end{document}